\documentclass{article}
\usepackage{amsmath,amssymb,amsthm}
\newtheorem{theorem}{Theorem}
\newtheorem{proposition}{Proposition}

\newtheorem{lemma}{Lemma}
\theoremstyle{definition}
\newtheorem{defn}{Definition}
\newcommand{\natnum}{\mathbb{N}}
\newcommand{\integers}{\mathbb{Z}}
\newcommand{\ratfield}{\mathbb{Q}}
\newcommand{\algnum}{\overline{\ratfield}}
\newcommand{\realfield}{\mathbb{R}}
\newcommand{\complexfield}{\mathbb{C}}
\newcommand{\comp}{\mathcal{C}}
\newcommand{\ptime}{\mathcal{P}}
\newcommand{\ptcreal}{\realfield_{\ptime}}
\newcommand{\ptccomplex}{\complexfield_{\ptime}}
\newcommand{\compreal}{\realfield_{\comp}}
\newcommand{\compcomplex}{\complexfield_{\comp}}

\title{On Polynomial Time Computable Numbers}
\author{Matsui, Tetsushi}
\date{}
\begin{document}
\maketitle
\begin{abstract}
  It will be shown that the polynomial time computable numbers form a field, and especially an algebraically closed field.
\end{abstract}
\section{Introduction}
\label{sec:intro}

The computable numbers have already appeared in Turing's paper~\cite{turi1936} in which he defined the Turing machine.
Turing defined them in real numbers, the definition is naturally extended to complex numbers.
Rice showed that the whole complex computable numbers forms an algebraically closed field~\cite{rice1954}.
The field is strictly larger than the algebraic number fields \(\algnum\) since it contains transcendental numbers such as \(\pi\), but is strictly smaller than the complex field \(\complexfield\) since it has only countable cardinality.

On the other hand, the polynomial time computability was getting more importance as a characteristic property of feasibly computable functions in the theory of computation.
Our main subject, the polynomial time computable numbers, can be, thus, considered as a definition of the feasibly computable numbers.
However, the interests it attracts are few; there is only a paper of Ko~\cite{ko__1983}, in which he investigates several ways of defining the polynomial time computable numbers.

In the paper, we will follow the direction of Rice; we will investigate algebraic properties of the set of whole polynomial time computable numbers, and in fact it forms an algebraically closed field.
The main theorem we will show is the following:

\begin{theorem}[main theorem]
\label{thm:main}
The set of whole polynomial time computable numbers forms an algebraically closed field.
\end{theorem}

In the next section, we will clarify the definition of computable numbers and discuss about the relation with Ko's definition.
The fact that the whole polynomial time computable numbers forms a field shall be shown in section~\ref{sec:field}.
Then, section~\ref{sec:theorem} shall be the proof of the main theorem.
Finally in section~\ref{sec:transcendental}, we shall show the fact that the circle ratio \(\pi\) is a polynomial time conputable numbers to proove that the set of polynomial time computable numbers is a proper superset of the field of algebric numbers.


\section{Definition}
\label{sec:definition}

At first, we introduce some notations and recall well-known results.

\(\natnum\), \(\integers\), \(\ratfield\), \(\realfield\) and \(\complexfield\) denote the set of natural numbers, the set of integers, the rational field, the real field and the complex field, respectively.

\subsection{Theory of Computation}
\label{sec:def_comp}

The first things we recall are terminologies in theory of computation.

\begin{defn}
\label{dfn:compfunc}
We call a function a {\em computable} function if there exists a deterministic Turing machine which terminates with the result of the function for each input.
\end{defn}

\begin{defn}
\label{dfn:ptcfunc}
We call a function a {\em polynomial time computable} function if there exists a deterministic Turing machine which terminates with the result of the function in time of a polynomial of the input size for each input.
\end{defn}

From now for simplicity, \(\comp\) denotes the class of computable functions, and \(\ptime\) the class of polynomial time computable functions.

Computable number means that it can be approximated by a computable function, and polynomial time computable number by a polynomial time computable function.
There are several ways to define them as definitions of real number: the Cauchy sequence, the Dedekind cut, etc.
We choose a way to define them through giving approximation fractions.
Accurately, the following is our definition.

\begin{defn}
A {\em computable number} is a complex number \(z\) that there exist two computable functions \(f\) and \(g\) from \(\natnum\) to \(\integers\) such that
\[\left|z - \frac{f(n)+g(n)i}{n}\right| \leq \frac{1}{n}\]
is satisfied for any natural number \(n\) greater than \(1\).
Especially, if \(z\) is a real number, we call it a {\em computable real number}.
\end{defn}

\begin{defn}
\label{dfn:ptcnum}
A {\em polynomial time computable number} is a complex number \(z\) that there exist two polynomial time computable functions \(f\) and \(g\) from \(\natnum\) to \(\integers\) such that
\[\left|z - \frac{f(n)+g(n)i}{n}\right| \leq \frac{1}{n}\]
is satisfied for any natural number \(n\) greater than \(1\).
Especially, if \(z\) is a real number, we call it a {\em polynomial time computable real number}.
\end{defn}

Let \(\compcomplex\) denote the set of whole computable numbers, and \(\ptccomplex\) the set of whole polynomial time computable numbers.
In case we restrict to real numbers \(\compreal\) and \(\ptcreal\) denote the set of computable real numbers and polynomial time computable real numbers respectively.

\subsubsection{Comparison with Ko's Definition}
\label{sec:cmp_ko}

We compare our definition of polynomial time computable real number and Ko's definition in~\cite{ko__1983}.

\begin{defn}[Ko's polynomial time computable real number]
\label{def:ko}
\(\phi:\natnum \rightarrow \{0,1\}^*\) {\em binary converge} to \(\alpha \in [0,1]\) means that
for any \(n \in \natnum\)
\[\log_2(\phi(n)) = n \land \left|\phi(n)2^{-n}-\alpha\right|\leq 2^{-n}\]
is satisfied.
\(\alpha\) is {\em polynomial time computable} if there is a computable function \(\phi\) binary converge to \(\alpha\) in time of a polynomial of \(n\).
\end{defn}

The most significant difference is that Ko defines it only in the range \([0,1]\).
Another point is that because Ko utilizes only \(2^n\) for denominators, the input \(n\) have to be interpreted unnaturally as if it is given by unary expansion.
Our definition uses functions which give numerators of approximation for any denominators.
We can show that the definitions are equivalent in the range \([0,1]\).
For the proof, we define the following sub-sequence of the natural numbers.

\begin{defn}
  \label{def:poly_incr_seq}
A sub-sequence of the natural numbers \(S\) is a {\em polynomially increasing sequence} if and only if there exist a polynomial time computable function \(\phi\) to enumerate each element \(s_i = \phi(i) \in S\) monotonically increasingly and a polynomial \(p\) satisfying \(s_i <s_{i+1} \leq p(s_i)\) for any \(i\).
\end{defn}

For example, an arithmetic sequence with a positive integer difference or a geometric sequence with a positive integer ratio are polynomially increasing sequences.
Especially, the sequence \(\{2^i\}\) appearing in Ko's definition is a polynomially increasing sequence.

\begin{lemma}
  \label{lem:partial}
A complex number \(z\) is a polynomial time computable number
if there exist polynomial time computable functions \(\hat{f},\hat{g}\)
from a polynomially increasing sequence \(S\) to \(\integers\) satisfying
\[\left|z - \frac{\hat{f}(n)+\hat{g}(n)i}{n}\right| \leq \frac{1}{n}\]
for any \(n \in S\).
\end{lemma}
\begin{proof}
We prove the lemma by constructing polynomial time computable functions \(f, g\) to define \(z\).

At first, for an element \(s\) of \(S\) we let \(f(s)=\hat{f}(s)\) and \(g(s)=\hat{g}(s)\).
For any \(m \in \natnum \setminus S\), we choose \(n \in S\) satisfying \((2+\sqrt{2})m\leq n\).
The size of \(n\) can be estimated as \(n < p((2+\sqrt{2})m)\) by the definition of \(S\).
We let \(f,g\) be \(f(m)=\lfloor\frac{\hat{f}(n)m}{n}\rceil\),  \(g(m)=\lfloor\frac{\hat{g}(n)m}{n}\rceil\), then \(f\) and \(g\) are polynomial time computable functions defining \(z\).
Actually,
\[f(m)=\left\lfloor\frac{\hat{f}(n)m}{n}\right\rceil = \frac{\hat{f}(n)m + \delta_1}{n}\]
\[g(m)=\left\lfloor\frac{\hat{g}(n)m}{n}\right\rceil = \frac{\hat{g}(n)m + \delta_2}{n}\]
by letting \(\delta_1, \delta_2\) denote adjustment terms for the roundings.
Then,
\[|\delta_i| \leq \frac{n}{2}\text{ (\(i=1\),\(2\))}\]
is satisfied and the total error is estimated as the following.
\begin{eqnarray*}
&&  \left|z - \frac{\left\lfloor\frac{\hat{f}(n)m}{n}\right\rceil + \left\lfloor\frac{\hat{g}(n)m}{n}\right\rceil i}{m}\right|\\
&=& \left|z - \frac{\frac{\hat{f}(n)m}{n} + \frac{\hat{g}(n)m}{n} i}{m} + \frac{\delta_1+\delta_2 i}{nm}\right|\\
&\leq& \frac{1}{n} + \left|\frac{1+i}{2m}\right| = \frac{1}{n} + \frac{1}{\sqrt{2}m}\\
&\leq& \frac{1}{(2+\sqrt{2})m} + \frac{1}{\sqrt{2}m}\\
&=& \frac{\sqrt{2}-1}{\sqrt{2}m} + \frac{1}{\sqrt{2}m} = \frac{1}{m}
\end{eqnarray*}

The functions \(f\) and \(g\) are polynomial time computable, because of the estimation \(n < p((2+\sqrt{2})m)\) already mentioned.
\end{proof}

The following proposition is a direct consequence of the lemma.
\begin{proposition}
  \label{prop:equivalent}
Ko's definition of polynomial time computable real numbers is equivalent to the restriction of our definition of polynomial time computable numbers in the range \([0, 1]\).
\end{proposition}

\subsubsection{Another Equivalent Definition}
\label{sec:other_def}

To use later in proofs, we prepare another form of definition of polynomial time computable numbers as a lemma.

\begin{lemma}
  \label{lem:rat_def}
A complex number \(z\) is a polynomial time computable number
if there exist two polynomial time computable functions \(F\) and \(G\)
from \(\natnum\) to \(\ratfield\) satisfying
\[\left|z - (F(n)+G(n)i)\right| \leq \frac{1}{n}\]
for any natural number \(n\).
\end{lemma}
\begin{proof}
We assume the existence of \(F\) an \(G\) satisfying the condition.
It is sufficient to show that the denominators for \(n\), which are not necessarily \(n\) for \(F(n)\) or \(G(n)\), can be \(n\).

We let \(f(n) = \left\lfloor nF(4n)\right\rceil\), \(g(n)=\left\lfloor nG(4n)\right\rceil\), and assume \(F(4n) = \frac{p_1}{q_1}\), \(G(4n) = \frac{p_2}{q_2}\).
Then, there exist \(d_i\) such that
\begin{eqnarray*}
f(n) &=& \frac{p_1n + d_1}{q_1}\\
g(n) &=& \frac{p_2n + d_2}{q_2}
\end{eqnarray*}
and \(d_i\) satisfy \(|d_i| \leq \frac{q_i}{2}\).
\begin{eqnarray*}
  \left|z - \frac{f(n)+g(n)i}{n}\right| &=& \left|z - \frac{p_1n+d_1}{q_1n}+\frac{(p_2n+d_2)}{q_2n}i\right|\\
&=& \left|z - (F(4n)+G(4n)i)-\frac{d_1}{q_1n}-\frac{d_2}{q_2n}i\right|\\
&\leq& \left|z - (F(4n)+G(4n)i)\right| + \left|\frac{d_1}{q_1n} + \frac{d_2}{q_2n}i\right|\\
&\leq& \frac{1}{4n} + \left|\frac{1}{2n} + \frac{1}{2n}i\right|\\
&=& \frac{1}{4n} + \frac{1}{\sqrt{2}n}\\
&<& \frac{1}{n}
\end{eqnarray*}

Since it is obvious that \(f\) and \(g\) are polynomial time computable functions, \(f\) and \(g\) satisfy the conditions of the definition~\ref{dfn:ptcnum}.
\end{proof}

\subsection{Algebra}
\label{sec:def_alg}

Finally, let's recall the definition of the algebraically closed fields.

\begin{defn}
\label{def:algclose}
An {\em algebraically closed field} is a field  \(K\) satisfying one of the following equivalent conditions.
  \begin{enumerate}
  \item Each element of \(K[X]\) has at least one root in \(K\).
  \item Each element of \(K[X]\) can be factored into linear factors in \(K[X]\).
  \end{enumerate}
\end{defn}

Popular examples of algebraically closed fields are the field of algebraic numbers \(\algnum\) and the complex number field \(\complexfield\).
As already mentioned, it is shown by Rice that the whole computable numbers form an algebraically closed field.


\section{The Field of Polynomial Time Computable Numbers}
\label{sec:field}

In this section, we show that the whole polynomial time computable numbers \(\ptccomplex\) is a field.
For ease at the beginning, we start from proving that the \(\ptcreal\) is a field.

\begin{lemma}
\label{lem:ptcreal_field}
The \(\ptcreal\) is a subfield of the \(\compreal\).
\end{lemma}
\begin{proof}
Since it is clear that the \(\ptcreal\) is a subset of the \(\compreal\), it only needs to show the lemma that the \(\ptcreal\) is closed under the four arithmetic operations: additions, multiplications, negations and inversions.

Let \(\lambda\) and \(\mu\) be elements of \(\ptcreal\).
By definition, there exist polynomial time computable functions \(f\) and \(g\) from \(\natnum\) to \(\integers\), and they satisfies:
\[\left|\lambda - \frac{f(n)}{n}\right| \leq \frac{1}{n}\]
and
\[\left|\mu - \frac{g(n)}{n}\right| \leq \frac{1}{n}\]
each for any \(n \geq 1\).

The first thing we show is the closedness under additions, i.e. \(\lambda + \mu\) is an element of the \(\ptcreal\).
By defining a function \(S\) from the natural numbers to the rationals by
\[S(n) = \frac{f(2n)+g(2n)}{2n},\]
it is a direct consequence of the fact that \(f\) and \(g\) are polynomial time computable functions that \(S\) is a function of polynomial time of input size \(\log n\) for the input \(n\).
\begin{eqnarray*}
&&  \left|\lambda+\mu - S(n)\right|\\
&=& \left|\lambda+\mu - \frac{f(2n)+g(2n)}{2n}\right|\\
&=& \left|\lambda - \frac{f(2n)}{2n}+\mu-\frac{g(2n)}{2n}\right|\\
&\leq& \left|\lambda - \frac{f(2n)}{2n}\right|+\left|\mu-\frac{g(2n)}{2n}\right|\\
&\leq& 2\cdot\frac{1}{2n} = \frac{1}{n}
\end{eqnarray*}
Therefore, \(\lambda + \mu\) has a defining polynomial time computable function from \(\natnum\) to \(\ratfield\), and it is a polynomial time computable real number from the lemma~\ref{lem:rat_def}. 
Thus, \(\ptcreal\) is closed under additions.

It is trivial to show the closedness under negation, or flipping the sign.

The next thing to show is the closedness under multiplications.
Let \(P\) be a function from \(\natnum\) to \(\ratfield\) s.t.
\[P(n) = \frac{f(cn)g(cn)}{c^2n^2},\]
where \(c\) is a constant \(|f(1)|+|g(1)|+4\).
It is clear that \(P\) is polynomial time computable.
\begin{eqnarray*}
    && \left|\lambda\mu - P(n)\right|\\
   &=& \left|\lambda\mu - \frac{f(cn)g(cn)}{c^2n^2}\right|\\
   &=& \left|\left(\lambda-\frac{f(cn)}{cn}\right)\left(\mu - \frac{g(cn)}{cn}\right)
     +\frac{f(cn)}{cn}\left(\mu - \frac{g(cn)}{cn}\right)
    +\frac{g(cn)}{cn}\left(\lambda-\frac{f(cn)}{cn}\right)\right|\\
&\leq& \left|\lambda-\frac{f(cn)}{cn}\right|\left|\mu - \frac{g(cn)}{cn}\right|
     +\left|\frac{f(cn)}{cn}\right|\left|\mu - \frac{g(cn)}{cn}\right|
     +\left|\frac{g(cn)}{cn}\right|\left|\lambda-\frac{f(cn)}{cn}\right|\\
&\leq& \frac{1}{c^2n^2}\
     +\left|\frac{f(cn)}{cn}\right|\frac{1}{cn}
     +\left|\frac{g(cn)}{cn}\right|\frac{1}{cn}\\
\end{eqnarray*}
Here, \(f\) satisfies \(\left|\frac{f(cn)}{cn}\right| \leq \left|f(1)\right|+\frac{3}{2}\) and so \(g\) does.
\begin{eqnarray*}
&\leq&\frac{1}{c^2n^2}\left(1 + \left(cn\left|f(1)\right|+\frac{3cn}{2}\right) + \left(cn\left|g(1)\right|+\frac{3cn}{2}\right)\right)\\
   &=&\frac{cn}{c^2n^2}\left(\left|f(1)\right| + \left|g(1)\right| +3 + \frac{1}{cn}\right)\\
&\leq&\frac{c^2n}{c^2n^2} = \frac{1}{n}
\end{eqnarray*}
Therefore, \(P\) is a polynomial time computable function from \(\natnum\) to \(\ratfield\) defining \(\lambda\mu\).
Again by lemma~\ref{lem:rat_def}, \(\lambda\mu\) is a polynomial time computable real number, and \(\ptcreal\) is closed under multiplications.

At last, we show the closedness under inversions.
Let \(\lambda\) be a polynomial time computable real number, which is not \(0\).
There exists a natural number \(k\) such that \(|f(k)|>1\), since \(\lambda\) is not \(0\).
With defining a polynomial \(p\) with the \(k\) as
\[p(X) = 2k^2X + k,\]
we define \(I\) a polynomial time computable function from \(\natnum\) to \(\ratfield\) by
\[I(n) = \frac{p(n)}{f(p(n))}.\]
Then,
\begin{eqnarray*}
    && \left|\lambda^{-1} - I(n)\right|\\
   &=& \left|\lambda^{-1} - \frac{p(n)}{f(p(n))}\right|\\
   &=& \left|\lambda^{-1}\right|\left|\frac{p(n)}{f(p(n))}\right|\left|\frac{f(p(n))}{p(n)} - \lambda\right|\\
&\leq& \left|\lambda^{-1}\right|\left|\frac{p(n)}{f(p(n))}\right|\frac{1}{p(n)}\\
&\leq& \left(\frac{|f(k)|}{k}-\frac{1}{k}-\frac{1}{p(n)}\right)^{-2}\frac{1}{p(n)}\\
   &=& \frac{k^2p(n)}{\left(p(n)\left(|f(k)|-1\right)-k\right)^2}
\end{eqnarray*}
Here, \(|f(k)| - 1 \geq 1\) holds.
\begin{eqnarray*}
 &\leq& \frac{k^2p(n)}{\left(p(n)-k\right)^2}\\
 &=& \frac{2k^4n+k^3}{(2k^2n)^2}\\
 &\leq& \frac{3k^4n}{4k^4n^2}\\
 &<& \frac{1}{n}
\end{eqnarray*}
Therefore, \(I\) is a polynomial time computable function from \(\natnum\) to \(\ratfield\) defining \(\lambda^{-1}\).
Once again by lemma~\ref{lem:rat_def}, \(\lambda^{-1}\) is a polynomial time computable real number, and \(\ptcreal\) is closed under inversion.
\end{proof}

By the lemma above, \(\ptcreal\) is a field.
The next lemma clarifies the relationship between \(\ptcreal\) and \(\ptccomplex\).

\begin{lemma}
\label{lem:ptcreal_ptccomplex}
A complex number \(z=x+yi\) is an element of \(\ptccomplex\) if and only if both its real part \(x\) and its imaginary part \(y\) are elements of \(\ptcreal\).
\end{lemma}
\begin{proof}
\(z=x+yi \in \ptccomplex\) clearly implies to \(x \in \ptcreal \land y \in \ptcreal\) by the definition.

Conversely, we assume  \(x \in \ptcreal \land y \in \ptcreal\).
Let \(f\) (\(g\)) be a defining function of \(x\) (\(y\) resp.).
Moreover, we define two functions \(\xi\) and \(\eta\):
\[\xi(n)=\frac{f(3n)+\kappa_1}{3}\]
\[\eta(n)=\frac{g(3n)+\kappa_2}{3},\]
where \(\kappa_1\) and \(\kappa_2\) are adjustment terms to keep the values of \(\xi\) and \(\eta\) respectively in integers, and thus their absolute values are at most \(1\).
Then,
\begin{eqnarray*}
&&  \left|x+yi -\frac{\xi(n)+\eta(n)i}{n}\right|\\
&=& \left|x+yi -\frac{f(3n)+g(3n)i+\kappa_1+\kappa_2i}{3n}\right|\\
&=& \sqrt{\left(x -\frac{f(3n)+\kappa_1}{3n}\right)^2+\left(y-\frac{g(3n)+\kappa_2}{3n}\right)^2}\\
&\leq& \frac{1}{3n}\sqrt{1 - 2\kappa_1(3nx-f(3n)) + \kappa_1^2 + 1 - 2\kappa_2(3ny-g(3n)) + \kappa_2^2}\\
&\leq& \frac{1}{3n}\sqrt{(\kappa_1 + 1)^2 + (\kappa_2 + 1)^2}\\
&\leq& \frac{\sqrt{8}}{3n}\\
&<& \frac{1}{n}
\end{eqnarray*}
Therefore, \(z=x+yi\) belongs to \(\ptccomplex\). 
\end{proof}

The two lemmas above imply the following theorem.

\begin{theorem}
  \label{thm:ptcfield}
The whole set of polynomial time computable numbers \(\ptccomplex\) forms a field.
\end{theorem}
\begin{proof}
From the lemma~\ref{lem:ptcreal_ptccomplex}, both real and imaginary parts of an element of \(\ptccomplex\) are elements of \(\ptcreal\).
All of four arithmetic operations of \(\ptccomplex\) are defined only from the four arithmetic operations of real and imaginary parts.
By the lemma~\ref{lem:ptcreal_field}, \(\ptcreal\) is closed under the four arithmetic operations.
Thus, both real and imaginary parts of the result of operations in \(\ptccomplex\) are in \(\ptcreal\).
From the lemma~\ref{lem:ptcreal_ptccomplex} again, the result is in \(\ptccomplex\).
It means that \(\ptccomplex\) is closed under all of four operations, and \(\ptccomplex\) forms a field.
\end{proof}


\section{Main Theorem}
\label{sec:theorem}

We will, in this section, prove the main theorem already stated.

\setcounter{theorem}{0}
\begin{theorem}
The set of whole polynomial time computable numbers \(\ptccomplex\) forms an algebraically closed field.
\end{theorem}
\setcounter{theorem}{2}

In the previous section, we have already shown that \(\ptccomplex\) is a field, and it is sufficient to show the algebraically closedness of the field.
By the definition~\ref{def:algclose}, it is the subject to prove that any \(\ptccomplex\)~coefficient polynomials have a root in \(\ptccomplex\) or they are factored into linear \(\ptccomplex\)~coefficient factors.
Note that existence of a root or factorization into linear factors are sufficient to prove, and it is not necessary to show that any given polynomial will be factored into linear factors {\em in polynomial time}.

Moreover, we can exclude polynomial with double roots transcendentally.
When one thinks about algebraic extensions, all roots including double roots can be constructed from single roots.

An outline of the proof will be as follows.
Think about all roots of given \(\ptccomplex\)~coefficient polynomials \(f\).
There are two factors of the errors in computation of the roots of \(f\).
The first factor is from expressing the coefficients in finite precisions.
The second factor is from terminating an algorithm of root finding at some precision.
It is, therefore, sufficient to show the theorem that making the errors from both factors in a desired precision takes at most a polynomial time of the precision.
We will show the following lemma for the approximations of coefficients.

\begin{lemma}
\label{lem:coeff_err}
Let \(f\) be a given \(\ptccomplex\)~coefficient monic polynomial.
For any natural number \(m\), it takes at most polynomial time of \(\log m\) to compute all coefficients of an approximation polynomial\footnote{We call a polynomial \(\tilde{f}\) an approximation polynomial of \(f\) if all coefficients of \(\tilde{f}\) are obtained from computing the defining function of the coefficients of \(f\).}  \(\tilde{f}\) of \(f\) so that roots \(\{\rho_i\}\) of \(f\) and \(\{\tilde{\rho}_i\}\) of \(\tilde{f}\) satisfy
\[\left|\rho_i - \tilde{\rho}_i\right|\leq \frac{1}{m}\]
for any \(i\) if appropriately arranged.
\end{lemma}

For the root approximations, we need two more lemmas.
Before stating the lemmas, let us name the condition that appears in the lemmas.
\begin{defn}
\label{def:conv_ini}
We call the following condition of a complex number \(\zeta\) for a polynomial \(f\) {\em converging initial condition}.
\begin{itemize}
\item There is an open convex set \(D\) to which \(\zeta\) belongs, and there exists a real number \(L>0\) such that for any \(z_1, z_2 \in D\) it satisfies
\[|f'(z_1)-f'(z_2)|\leq L|z_1 - z_2|.\]
\item \(f'(\zeta) \neq 0\) holds, and there are real numbers \(a, b\) such that \(|f'(\zeta)^{-1}|\leq a\), \(|f'(\zeta)^{-1}f(\zeta)|\leq b\) and \(h = abL \leq \frac{1}{2}\).
\item A closed disc \(\overline{U}\) with radius \(t^{*}=(1-\sqrt{1-2h})/(aL)\) centered at \(\zeta\) or:
\[\overline{U}=\{z\in\complexfield ; |z-\zeta|\leq t^{*}\}\]
is in \(D\).
\end{itemize}
\end{defn}

The two lemmas follow.

\begin{lemma}
  \label{lem:newton_method}
Let \(f\) be a \(\ratfield[i]\)~coefficient monic polynomial, which has no multiple roots.
If \(\rho^{(0)}\) satisfying converging initial condition for \(f\) is given,
then for any natural number \(m\), it takes at most polynomial time of \(\log m\) to compute an approximation \(\tilde{\rho}\) of the root \(\rho\) of f to satisfy
\[\left|\rho - \tilde{\rho}\right|\leq \frac{1}{m}.\]
\end{lemma}

\begin{lemma}
 \label{lem:common_initial}
Let \(f_1, f_2, \ldots\) be a series of polynomials uniformely converging to a polynomial \(f \in \complexfield[X]\) with no double roots in any compact region..
Then, there exists \(m_0\) such that there exists \(\rho^{(0)}\) which satisfies converging initial condition for any \(f_m\) (\(m>m_0\)).
\end{lemma}

\subsection{Proof of the Main Theorem}
\label{sec:proof_main}

Firstly, we prove the main theorem assuming the validity of the lemmas.

\begin{proof}[proof of the theorem~\ref{thm:main}]
We prove the theorem by showing that each \(\ptccomplex\)~coefficient polynomial has at least one root in \(\ptccomplex\).

The case of degree \(1\) is trivial, thus we assume that \(f\) is a polynomial of degree \(d > 1\).
Without loss of generality, we can assume that \(f\) is monic.
Moreover, we can assume that \(f\) is irreducible over a field over \(\ratfield\) generated by all coefficients of \(f\).
Otherwise, \(f\) can be factored into lower degree irreducible polynomials.

Then, by the lemma~\ref{lem:coeff_err}, for any \(m\), an approximation polynomial \(\tilde{f}_m\), whose roots are distant at most \(\frac{1}{2m}\) from each of the roots of \(f\), can be computed in polynomial time of \(\log 2m\), and thus of \(\log m\).

Since the approximation polynomials \(\{\tilde{f}_m\}_{m=1}^{\infty}\) of \(f\) uniformely converge to \(f\) in any compact regions of \(\complexfield\),
there exists a number \(m_0\) such that there exists \(\rho^{(0)}\) which satisfies converging initial condition for any \(\tilde{f}_m\) (\(m>m_0\)) by the lemma~\ref{lem:common_initial}.
By the lemma~\ref{lem:newton_method}, then, we can compute an approximation of a root of \(\tilde{f}_m\) in precision of \(\frac{1}{2m}\) from \(\rho^{(0)}\) in polynomial time of \(\log 2m\) and thus of \(\log m\) again.
The cases of \(m \leq m_0\) are ignorable since they are only finite numbers.

As a consequence, a root of \(f\) can be computed for any natural number \(m\)  in polynomial time of \(\log m\) with an error at most \(\frac{1}{m}\).
Therefore, it is an element of \(\ptccomplex\).
\end{proof}

The rest of the section consists of proofs of the lemmas.

\subsection{Proof of the Lemma~\ref{lem:coeff_err}}
\label{sec:proof_lem_coeff_err}

In this section, we prove the lemma~\ref{lem:coeff_err}.
In the proof, the following theorem plays the central role.

\begin{theorem}[Ostrowski]~\cite[section 2.3]{hous1970}
  \label{thm:coef_err}
Let \(f\) and \(g\) be two different complex coefficient polynomials.
\begin{eqnarray*}
  f(X) &=& X^n + a_{n-1}X^{n-1} + \cdots + a_1X + a_0\\
  g(X) &=& X^n + b_{n-1}X^{n-1} + \cdots + b_1X + b_0\\
\end{eqnarray*}
Moreover, \(\gamma\) denotes a real constant:
\[\gamma = 2 \max (\{|a_{n-j}|^{1/j}\}\cup \{|b_{n-j}|^{1/j}\})\]
and \(\epsilon\) denotes a positive real number satisfies:
\[\epsilon^n = \sum_{j=0}^{n-1}|b_j - a_j|\gamma^j.\]

Then, zeros \(z_k\) (\(w_k\)) of \(f\) (\(g\) resp.) can be arranged to satisfy
\[|z_j - w_j|<2n\epsilon.\]
\end{theorem}

\begin{proof}[proof of the lemma~\ref{lem:coeff_err}]
We write the given polynomial \(f\) explicitly:
\[f(X)= X^n + a_{n-1}X^{n-1} + \cdots + a_1X + a_0\]
with \(a_i\) in \(\ptccomplex\).
Let \(\tilde{f}\)
\[\tilde{f}(X)= X^n + \tilde{a}_{n-1}X^{n-1} + \cdots + \tilde{a}_1X +\tilde{a}_0\]
be an approximation polynomial of \(f\).

Let \(\gamma_{f,\tilde{f}}\) and \(\epsilon_{f,\tilde{f}}\) be:
\begin{eqnarray*}
  \gamma_{f,\tilde{f}} &=& 2 \max (\{|a_{n-j}|^{1/j}\}\cup \{|\tilde{a}_{n-j}|^{1/j}\})\\
  \epsilon_{f,\tilde{f}}^n &=& \sum_{j=0}^{n-1}|\tilde{a}_j - a_j|\gamma_{f,\tilde{f}}^j.
\end{eqnarray*}
Then, by the theorem~\ref{thm:coef_err}, the roots \(\{\rho_i\}\) of \(f\) and \(\{\tilde{\rho}_i\}\) of \(\tilde{f}\) satisfy the following inequality:
\[\left|\rho_i - \tilde{\rho}_i\right|<2n\epsilon_{f,\tilde{f}}\]
with an appropriate arrangement.

At first, we estimate \(\gamma_{f,\tilde{f}}\).
Since \(\tilde{a}_{n-j}\) is calculated from the defining function of \(a_{n-j}\), \(|\tilde{a}_{n-j}|<|a_{n-j}|+1\) holds.
Then,
\[\gamma_{f,\tilde{f}} < 2\max\{(|a_{n-j}|+1)^{1/j}\}\]
and by letting \(\gamma\) denote the right hand side,
\(\gamma \geq 1\) holds.

Secondly, we estimate \(\epsilon_{f,\tilde{f}}\).
By replacing \(\gamma_{f,\tilde{f}}\) in the estimation of \(\epsilon_{f,\tilde{f}}\) by \(\gamma\), it holds that:
\[\epsilon_{f,\tilde{f}}^n = \sum_{j=0}^{n-1}|\tilde{a}_j - a_j|\gamma_{f,\tilde{f}}^j < \sum_{j=0}^{n-1}|\tilde{a}_j - a_j|\gamma^j.\]
If for any \(j\)
\[|\tilde{a}_j - a_j|\gamma^j \leq \frac{1}{k}\]
hold, then
\[\epsilon_{f,\tilde{f}} < \sqrt[n]{\frac{n}{k}}\]
is implied.

To make the all differences of the roots be smaller than \(\frac{1}{m}\),
using the theorem
\[\left|\rho_i - \tilde{\rho}_i\right|<2n\epsilon_{f,\tilde{f}}\]
and the assumption above, we can obtain:
\begin{eqnarray*}
2n\epsilon_{f,\tilde{f}} < 2n\sqrt[n]{\frac{n}{k}} &\leq& \frac{1}{m}\\
k^{\frac{1}{n}} &\geq& 2n^{1+\frac{1}{n}}m\\
k &\geq& 2^n n^{n+1} m^n.
\end{eqnarray*}
Therefore, it is sufficient if computation is carried out to the place
\(|\tilde{a}_j - a_j|\gamma^j\) is smaller than \((2^n n^{n+1} m^n)^{-1}\).
The errors in coefficients themselves are estimated as
\[|\tilde{a}_j - a_j| \leq \frac{1}{k\gamma^j}\leq \frac{1}{2^nn^{n+1}m^n\gamma^j}.\]
Since \(a_j\) of a coefficient of \(f\) is a polynomial time computable number,
it takes at most polynomial steps of \(\log (2^n n^{n+1}  \gamma^j m^n)\) to calculate \(\tilde{a}_j\) with error at most \((2^n n^{n+1} \gamma^j m^n)^{-1}\) .
Because \(n\) and \(\gamma\) are constant depending only on \(f\), the total time complexity is a polynomial of \(\log m\).
For all other coefficients, the estimations are similar, thus the total complexity is a sum of \(n\) polynomials, i.e. the total complexity is polynomial time.
\end{proof}

\subsection{Proof of the Lemma~\ref{lem:newton_method}}
\label{sec:proof_newton_method}

In this section, we prove the lemma~\ref{lem:newton_method}.
In the proof, the following theorem is used.

\begin{theorem}[Kantorovich]~\cite[section 4.3]{sm__1994}
  \label{thm:convergence}
Let \(f(\mathbf{x})\) be a differentiable function defined in a open convex set \(D\) of \(\realfield^n\), \(\mathbf{x}^{(0)}\) be in \(D\).
Assume the following conditions are satisfied.
\begin{itemize}
\item Jacobian \(J(\mathbf{x})\) is Lipschitz continuous on \(D\),
i.e.\ it satisfies:
\[||J(\mathbf{x})-J(\mathbf{y})||\leq L||\mathbf{x}-\mathbf{y}||\]
where \(\mathbf{x}\), \(\mathbf{y} \in D\) and \(L>0\).
\item \(J(\mathbf{x}^{(0)})\) is regular and it satisfies \(||J(\mathbf{x}^{(0)})^{-1}||\leq a\), \(||J(\mathbf{x}^{(0)})^{-1}f(\mathbf{x}^{(0)})||\leq b\) and \(h = abL \leq 1/2\).
\item An open ball \(\overline{U}\) centered at \(\mathbf{x}^{(0)}\) and with the diameter \(t^{*}=(a-\sqrt{1-2h})/(aL)\)
\[\overline{U}=\{\mathbf{x}\in\realfield^n\mid ||\mathbf{x}-\mathbf{x}^{(0)}||\leq t^{*}\}\]
is in \(D\).
\end{itemize}
Then,
\begin{itemize}
\item There exists in \(\overline{U}\) only a solution \(\mathbf{x}^{*}\) of \(f(\mathbf{x})=\mathbf{0}\).
\item An approximation sequence of the Newton method \(\{\mathbf{x}^{(\nu)}\}\) with the initial value \(\mathbf{x}^{(0)}\) is defined, and then \(\mathbf{x}^{(\nu)} \in D\) and
\[||\mathbf{x}^{(\nu)}-\mathbf{x}^{*}||\leq \frac{(1-\sqrt{1-2h})^{2^{\nu}}}{2^{\nu}aL}\]
hold for any \(\nu\).
\end{itemize}
\end{theorem}

\begin{proof}[proof of the lemma~\ref{lem:newton_method}]
From the formula of the Newton method giving an approximation sequence \(\{\rho^{(\nu)}\}\) of a root starting from \(\rho^{(0)}\) is
\[\rho^{(\nu+1)}=\rho^{(\nu)}-\frac{f(\rho^{(\nu)})}{f'(\rho^{(\nu)})}\]
and it takes at most \(O(\deg f)\) rational operations per step.
Moreover, by the inequalty of the theorem, we can estimate the number of iterations \(\nu\) such that:
\[\frac{(1-\sqrt{1-2h})^{2^{\nu}}}{2^{\nu}aL} \leq \frac{1}{m}\]
in terms of \(m\).
The fact that \(h=\frac{1}{2}\) holds only if the approximating root is a multiple root implies \(h < \frac{1}{2}\) since \(f\) is assumed not to have a multiple root, thus the constant \(1-\sqrt{1-2h}\) of the numerator is less than \(1\).
The power of the numerator, therefore, dominates over the factor \(2^{\nu}\) of the denominator.
As a consequence, we can use the following inequality for large \(\nu\):
\[\frac{(1-\sqrt{1-2h})^{2^{\nu}}}{2^{\nu}aL}\leq \frac{(1-\sqrt{1-2h})^{2^{\nu}}}{2aL}.\]
Then,
\begin{eqnarray*}
  \frac{(1-\sqrt{1-2h})^{2^{\nu}}}{2aL} &\leq& \frac{1}{m}\\
  (1-\sqrt{1-2h})^{2^{\nu}} &\leq& \frac{2aL}{m}\\
  2^{\nu} \log (1-\sqrt{1-2h}) &\leq& \log(2aL) - \log m\\
\end{eqnarray*}
Let a constant \(c_0\) be
\[c_0=\frac{1}{-\log (1-\sqrt{1-2h})},\]
then,
\begin{eqnarray*}
  2^{\nu} &\geq&  c_0\log m - c_0\log(2aL)\\
  \nu \log 2 &\geq& \log \left(c_0\log m - c_0\log(2aL)\right)\\
&=& \log\log m + \log c_0 + \log\left(1 - \frac{\log(2aL)}{\log m}\right)\\
\nu &\geq& \frac{\log\log m}{\log 2} + \frac{\log c_0 + \log\left(1 - \frac{\log(2aL)}{\log m}\right)}{\log 2}.
\end{eqnarray*}
We can conclude that the time complexity to obtain an approximation \(\rho^{(\nu)}\) distant from a root \(\rho\) of \(f\) at most \(\frac{1}{m}\) starting from \(\rho^{(0)}\) is \(O(\log\log m)\) rational operations.
Thus, it is polynomial time.
\end{proof}

\subsection{Proof of Lemma~\ref{lem:common_initial}}

In this section, we prove the last lemma.

\label{sec:proof_common_initial}
\begin{proof}
Since the given \(f\) has no double roots and \(\{f_i\}\) converges to \(f\),
a definition
\[\mu_0 = \max(\{0\} \cup \{m \in \natnum; f_m \text{has double roots}\})\]
makes sense.

A root \(\rho\) of \(f\) is not a root of \(f'\), i.e.\ \(f'(\rho) \neq 0\), since again \(f\) has no double roots.
In the following proof, we will fix a root \(\rho\) of \(f\) and let \(w\) be \(|f'(\rho)|\).
Choose and fix an arbitrary real number \(\alpha\) in a range \(0 < \alpha < 1\).
Then, there are at most finitely many \(m\) satisfying
\(|f_m'(\rho)| < \alpha w\),
because \(\{f_i'(\rho)\}\) converge to \(f'(\rho)\).
Let 
\[\mu_1 = \max(\{\mu_0\} \cup \{m; |f_m'(\rho)| < \alpha w\})\]
and
\[A(\mu_1, \alpha) = \bigcup_{m>\mu_1}\{z \in \complexfield; |f_m'(z)|<\alpha w\}.\]
Then, \(\rho\) does not belong to \(A(\mu_1, \alpha)\) by the definition of \(\mu_1\).

The next step is to determine an open convex set \(D\) to which \(\rho\) belpngs.
Let \(\delta_A\) be the distance between \(\rho\) and \(A(\mu_1, \alpha)\):
\[\delta_A = \inf \{|\rho - z|; z \in A(\mu_1, \alpha)\}\]
and
\[D = D(\delta_A) = \{z \in \complexfield; |\rho - z| < \delta_A\}.\]
It is clear that \(f\) and all of \(\{f_i\}\) are Lipschitz continuous in \(D\), since they are polynomials.
Actually, \(L\) can be:
\[L = \max(\frac{1}{2}, \sup\{|f_m''(z)|; z \in D \land m > \mu_1\}).\]

To satisfy the rest of the conditions, let \(\delta_0 = \min(\delta_A, \frac{2\alpha w}{L})\) and let \(\gamma\) be \(1\) if \(\delta_0 \neq \frac{3\alpha w}{2L}\) or an arbitrarily chosen real number satisfying \(\frac{1}{2} \leq \gamma < 1\) otherwise.
Besides, we define \(\beta\) as:
\[\beta = \gamma \left(\frac{2\delta_0}{3w}-\frac{2L\delta_0^2}{9\alpha w^2}\right).\]
Then \(h=\alpha^{-1}\beta L < \frac{1}{2}\).
Actucally,
\begin{eqnarray*}
  h &=&  \alpha^{-1}\beta L\\
&=& \alpha^{-1} L \gamma \left(\frac{2\delta_0}{3w}-\frac{2L\delta_0^2}{9\alpha w^2}\right)\\
&=& \gamma \frac{2 L \delta_0}{3 \alpha w}\left(1 - \frac{L \delta_0}{3 \alpha w}\right)
\end{eqnarray*}
By writing \(\xi = \frac{L \delta_0}{3 \alpha w}\), we have:
\[  h = 2 \gamma \xi (1 - \xi)\]
and by solving \(h < \frac{1}{2}\) we obtain \(\xi \neq \frac{1}{2}\).
Therefore, if \(\delta_0 \neq \frac{3\alpha w}{2L}\) then \(h < \frac{1}{2}\) holds.
On the other hand, if \(\delta_0 = \frac{3\alpha w}{2L}\) then \(h = \frac{\gamma}{2}\) and by the definition of \(\gamma\) we have \(h < \frac{1}{2}\).

We think a set determined with the \(\beta\) and a parameter \(\mu\):
\[B(\mu, \beta) = \bigcup_{m>\mu} \{z \in \complexfield; |f_m'(z)^{-1}f_m(z)|>\beta w\}\]
and we would like to settle \(\mu\) so that \(\rho\) does not belong to \(B(\mu, \beta)\).
Transformation of the condition of each set consisting \(B(\mu, \beta)\) gives
\(|f_m(z)|>\beta w |f_m'(z)|\).
Since we know that \(\rho \not \in A(\mu_1, \alpha)\), by letting \(\mu \geq \mu_1\) it is sufficient for the condition to be satisfied outside \(A(\mu_1, \alpha)\).
A condition \(|f_m'(z)|\geq \alpha w\) is satisfied outside \(A(\mu_1, \alpha)\), thus
\[|f_m(z)|>\beta w |f_m'(z)| > \alpha \beta w^2.\]
Especially, at \(\rho\), there are at most finitely many \(m\) to satisfy \(|f_m(\rho)|> \alpha \beta w^2\) since \(\{f_i\}\) converge to \(f\).
Then, with
\[\mu_2 = \mu_2(\beta) = \max(\{\mu_1\} \cup \{m; |f_m(\rho)|> \alpha \beta w^2\}),\]
\(\rho\) does not belong to \(B(\mu_2, \beta)\).
We let \(\delta_B\) denote the distance between \(\rho\) and \(B(\mu_2, \beta)\):
\[\delta_B = \delta_B(\beta) = \inf\{|\rho - z|; z \in B(\mu_2, \beta)\}.\]

Finally, let \(\hat{\gamma}\) be \(1-\sqrt{1-\gamma}\) and
\[\delta = \min(\delta_B, \frac{\hat{\gamma}\delta_0}{3}).\]
Then, with a point \(\rho^{(0)}\) inside the disc centered at \(\rho\) with the radius \(\delta\), a closed disc \(\overline{U}\) centered at \(\rho^{(0)}\) with the radius \(\frac{2\hat{\gamma}\delta_0}{3}\) is contained in \(D\) and \(\rho\) belongs to \(\overline{U}\).

Conseqeuently, by choosing \(m_0\) as \(\mu_2\), \(\rho^{(0)}\) satisfies the converging initial condition for \(f_m\) for any \(m>m_0\).
\end{proof}



\section{Transcendental Numbers}
\label{sec:transcendental}

To show that the field of polynomial time computable numbers \(\ptccomplex\) contains a part of transcendental numbers, we demonstrate that the circle ratio \(\pi\) is a polynomial time computable number.

\begin{proposition}
\label{prop:pi_is_ptcn}
The circle ratio \(\pi\) is a polynomial time computable number.
\end{proposition}
\begin{proof}
There are numbers of methods to compute the circle ratio.
We choose Machin's formula:
\[\frac{\pi}{4} = 4\arctan\left(\frac{1}{5}\right) - \arctan\left(\frac{1}{239}\right).\]

Since a sum of polynomial time computable numbers is also a polynomial time computable number as shown in the lemma~\ref{lem:ptcreal_field}, it is sufficient to show that for any \(k > 1\), \(\arctan\left(\frac{1}{k}\right)\) is a polynomial time computable number.
The Taylor expansion of \(\arctan\) for \(|x|<1\) is:
\[\arctan(x) = \sum_{i=0}^{\infty}\frac{(-1)^i x^{2i+1}}{2i+1}\]
and, by truncating to \(m\) terms, the error is at most \(\frac{x^{2m+1}}{(2m+1)}\).
Ignoring the numerator for simplicity, \(m\) for at most \(\frac{1}{n}\) error can be estimated as the following.
\begin{eqnarray*}
  \left(\frac{1}{k}\right)^{2m+1} &<& n^{-1}\\
(2m+1)\log \left(\frac{1}{k}\right) &<& -\log n\\
2m+1 &>& \frac{\log n}{\log k}\\
m &>& \frac{\log n - 1}{2\log k}\\
\end{eqnarray*}
Thus the number of terms can be at most \(\log n\).

The computations of evaluating the \(m\) term expansion of \(\arctan\left(\frac{1}{k}\right)\) is estimated as follows.
We compute
\[\left|\arctan\left(\frac{1}{k}\right) - \sum_{i=0}^{m-1}\frac{(-1)^i}{(2i+1) k^{2i+1}}\right| < \frac{1}{n}\]
by using a common denominator:
\[\frac{1}{(2m-1)!!k^{2m-1}}\sum_{i=0}^{m-1}(-1)^i k^{2m-2-2i}\prod_{j=0}^{m-1-i}(2j+1).\]
Then the size of the denominator is
\begin{eqnarray*}
\log \left((2m-1)!!k^{2m-1}\right) &<& \log \left([(2m-1)k]^{2m-1}\right)\\
&=& (2m-1)(\log (2m-1) + \log k)\\
&=& O(\log n)O(\log\log n) = O(\log n \log\log n)
\end{eqnarray*}
and it requires about \(2m\) multiplications.
Thus, the number of steps is estimated as:
\[O(\log n)O((\log n \log\log n)^2) = O(\log^4 n).\]

Similarly, the number of steps to compute the numerator is \(O(\log^4 n)\).
Thus, the total number of steps is also \(O(\log^4 n)\).

Therefore, \(\arctan(\frac{1}{k})\) are polynomial time computable and the circle ratio \(\pi\) as a sum of them is a polynomial time computable number.
\end{proof}


\section{Concluding Remarks}
\label{sec:conclusion}

We showed that the field of polynomial time computable numbers \(\ptccomplex\) is an algebraically closed field.
Because there are transcendental numbers including \(\pi\) in \(\ptccomplex\), the field is a proper extension of \(\algnum\) the algebraic closure of \(\ratfield\).
On the other hand, because the class of polynomial time computable functions does not contain any EXPTIME-complete functions by the hierarchy theorem~\cite[section 7]{kasa1987}, the field is a proper subfield of the field of whole computable numbers \(\compcomplex\).
One may be interested in whether the subsets of \(\compcomplex\) corresponding to other complexity classes are also algebraically closed fields or not.
By following the argument of this paper, it is easy to conclude that such sets corresponding to any classes containing \(\ptime\) are algebraically closed fields.
We do not know about the proper subclasses of \(\ptime\), even whether the corresponding sets are fields or not.

\section*{Acknowledgment}
\label{acknowledgment}

We express sincere thanks to Prof. Dr. Nakamula Ken for his kind comments especially for letting me know the theorem of Ostrowski.

\bibliographystyle{plain}
\bibliography{computables}
\end{document}